\documentclass[english]{lni}

\IfFileExists{latin1.sty}{\usepackage{latin1}}{\usepackage{isolatin1}}

\usepackage[T1]{fontenc}
\usepackage[utf8]{inputenc}
\usepackage[usenames,dvipsnames]{color}
\usepackage{graphicx}
\usepackage{booktabs}
\usepackage{subfigure}
\usepackage{xspace}
\usepackage{url}

\newcommand{\corrwidth}{60mm}
\newcommand{\conceptwidth}{100mm}
\newcommand{\indexwidth}{70mm}
\newcommand{\tg}[1]{\textcolor{Orange}{Thomas: #1}}
\renewcommand{\tg}[1]{}

\newcommand{\code}[1]{\textsf{\small #1}\xspace}
\newcommand{\subtxt}[1]{\textit{\footnotesize #1}\xspace}

\definecolor{gray}{gray}{0.50}
\definecolor{darkgreen}{rgb}{0,0.6,0}

\newcommand{\tickYes}{{\color{darkgreen}$\surd$}\xspace}
\newcommand{\tickNo}{{\color{red}$\times$}\xspace}
\newcommand{\tickGood}{{\color{darkgreen}+}\xspace}
\newcommand{\tickOk}{{\color{gray}o}\xspace}
\newcommand{\tickBad}{{\color{red}--}\xspace}
\renewcommand{\code}[1]{\textsf{\small #1}\xspace}

\begin{document}

\pagestyle{headings}  % switches on printing of running heads

\title{Measuring the Accuracy of Linked Data Indices}

\author{
	Thomas Gottron \\
	\\
Institute for Web Science and Technologies\\
University of Koblenz-Landau\\
Universitätsstraße 1\\
56070 Koblenz\\
\url{gottron@uni-koblenz.de}
}

\maketitle

\begin{abstract}
Being based on Web technologies, Linked Data is distributed and decentralised in its nature.
Hence, for the purpose of finding relevant Linked Data on the Web, search indices play an important role.
Also for avoiding network communication overhead and latency, applications rely on indices or caches over Linked Data.
These indices and caches are based on local copies of the original data and, thereby, introduce redundancy.
Furthermore, as changes at the original Linked Data sources are not automatically propagated to the local copies, there is a risk of having inaccurate indices and caches due to outdated information.
In this paper I discuss and compare methods for measuring the accuracy of indices.
I will present different measures which have been used in related work and evaluate their advantages and disadvantages from a theoretic point of view as well as from a practical point of view by analysing their behaviour on real world data in an empirical experiment.
\end{abstract}

%*******************************************************************************
\section{Introduction}
\label{sec:introduction}

The Linked Open Data (LOD) movement implements Tim Berner-Lee's vision of a \emph{Web of Data}.
The principles underlying Linked Data enable data providers to model, interlink and publish their data on the Web in a distributed and decentralised way.
On the consumer's side the Web-oriented technological basis (i.e. using RDF and HTTP) permits to easily make use of the data and integrate it into various applications.
These benefits and advantages of the Linked Data idea lead to, on the one hand, more and more data providers to contribute to the Web of Data and, on the other hand, more and more developers to spawn new applications making use of this data.
However, the growth and conceptual nature of Linked Data poses some technological challenges.
The distributed nature of Linked Data calls for search engines and indices to provide data catalogues of what kind of data is available and where it can be found.
Furthermore, some applications use data caches to avoid communication overhead when accessing data on the Web.

\begin{figure}[btp]
\centering
	\includegraphics[width=\conceptwidth]{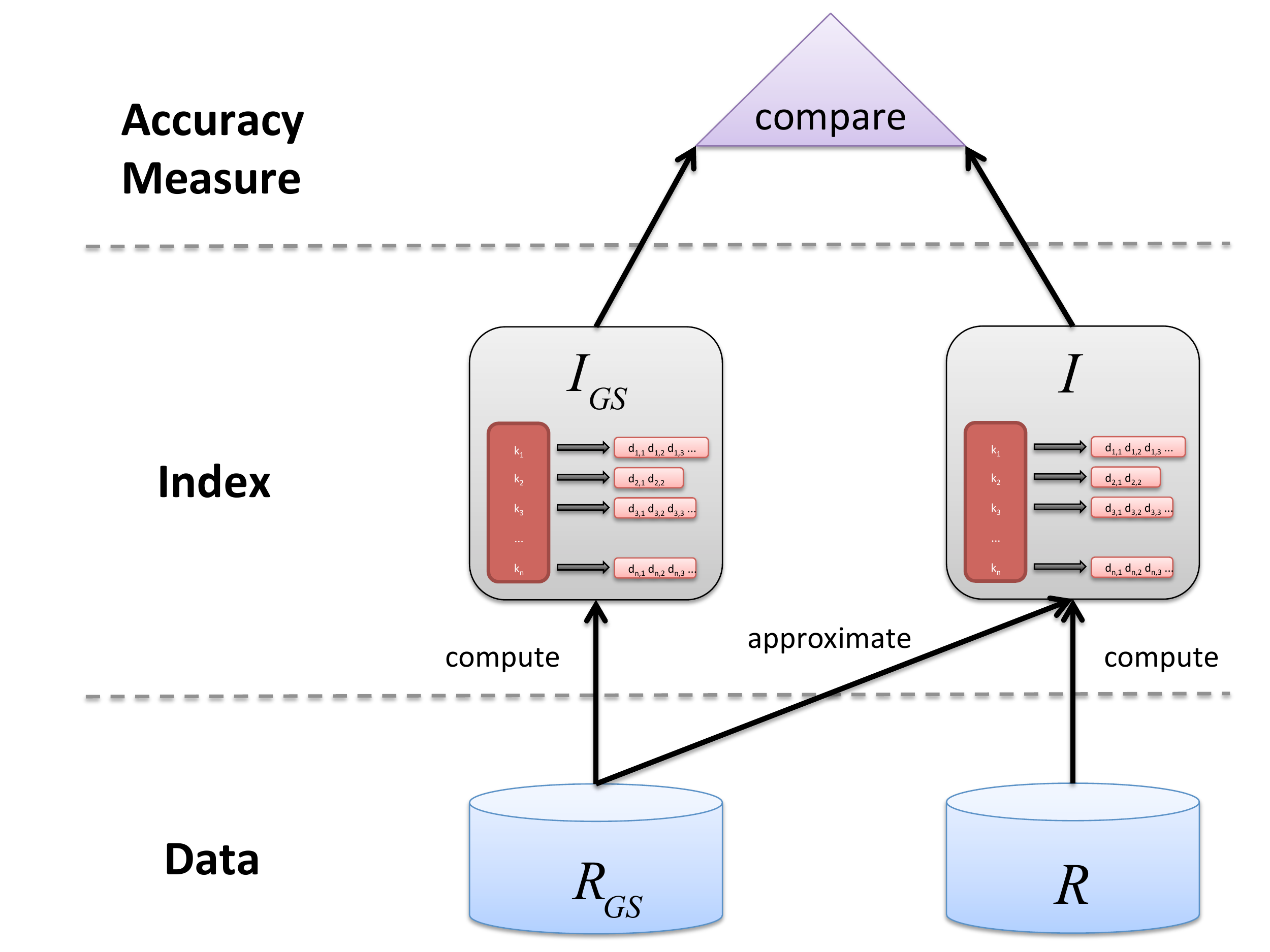} 
\caption{Inaccurate indices $I$ can occur when computed over outdated data $R$ or when computed in an approximative way over current data $R_{\textsc{gs}}$. A measure of accuracy needs to compare such an index $I$ to a perfect, gold standard index $I_{\textsc{gs}}$ computed over $R_{\textsc{gs}}$ in a lossless way.}
\label{fig:concept}
\end{figure}

Such indices and caches correspond to local views on the data.
These local views might be inaccurate in the sense that they do not reflect the state of the data at the original data sources. 
There are two main reasons for indices constituting an inaccurate view on the data.
The first reason is rooted in the decentralised and dynamic nature of Linked Data.
Changes at the original data sources are not automatically propagated to the applications and their indices and data caches.
Accordingly a task which needs to be addressed in the context of Linked Data applications is the active maintenance of indices.
This means, that applications have to synchronise their local view on the data with the data at the origin~\cite{P:LDOW:2010:UmbrichHHP,P:ESWC:2014:DividinoKG}.
A second reason are applications which build their indices only in an approximate way~\cite{J:JWS:2012:KonrathGSS}.
This happens, for instance, to achieve scalability and to be able to efficiently process large volumes of Linked Data.
Figure~\ref{fig:concept} illustrates these two paths which can lead from a Linked Data set $R_{\textsc{gs}}$ and an older version of this data set $R$ to an index $I$ which is inaccurate compared to a \emph{perfect} index $I_{\textsc{gs}}$.

In both cases, there is a certain tradeoff between index accuracy and the commitment of limited resources.
When updating and maintaining an index, the limited resource is typically network bandwidth.
Spending more effort on updating an index requires more bandwidth but will yield indices of higher accuracy.
Likewise, the approximate computation of indices can usually be influenced by a parameter which trades index accuracy for required computational resources.
Hence, given the limitations of available resources the providers of LOD indices will need to strike a balance between the effort for building and maintaining an index and the quality of service, i.e. the accuracy of their index.

While measuring the costs for computation and network bandwidth has been addressed in many other fields,  measures for the accuracy of a Linked Data index are less established.
In related work, different methods for measuring index accuracy have been described and applied.
But, rarely more than one approach is used. 
This makes it difficult to compare results.
Thus, the idea of this paper is to compare different measures for index accuracy on a theoretical and practical level.
This should help in judging and comparing existing results as well as to support the choice of measures to use in future research work in this field.
In more detail, this paper will make three contributions:

\begin{description}

\item[Survey of Measures:] Following a review of related work, this paper provides an extensive overview of approaches and methods used to measure the accuracy of index structures over Linked Data. 
All methods are presented in a unified and formal way for ease of comparison. 

\item[Theoretical Comparison:] Considering the general task of RDF indices and the specific Web based setting of Linked Data, this paper compares approaches and methods to measure the accuracy of LOD indices w.r.t their theoretical limitations and advantages.

\item[Practical Comparison:] By using an established data set of evolving Linked Data the paper analyses the approaches and methods to measure the accuracy of LOD indices w.r.t their behaviour in practice. 
In particular it presents an analysis of the correlation of the measures to understand how far similar or different notions of index accuracy are captured by the discussed measures.

\end{description}

As a next step we look at related work in Section~\ref{sec:related_work} to get an overview of the context in which LOD indices are used and evaluated for their accuracy. 
This section will also provide a first and brief review of the methods used to measure index accuracy.
Subsequently, Section~\ref{sec:index} will present an abstract and formal representation for Linked Data indices.
This formalisation allows to abstract from concrete implementations and serves as basis for the unified definition of index accuracy measures in Section~\ref{sec:measures}.
The theoretic analysis and comparison of the measures is presented in Section~\ref{sec:theory} and the practical and empirical comparison in Section~\ref{sec:empirical}.
The findings are discussed in~\ref{sec:discussion} and the paper is concluded with a summary and an outlook at future work.

% ****************************************************************************************
% ****************************************************************************************
%   Section break
% ****************************************************************************************
% ****************************************************************************************

\section{Related Work}
\label{sec:related_work}

In recent years, various index models over LOD have been proposed.
Many of them focus on specific aspects of the data or are dedicated to support application specific tasks.
When looking at the RDF basis of LOD, one has to consider also the works on indices for RDF triple stores, such as Hexastore~\cite{J:VLDB:2008:WeissKB} or RDF3X~\cite{J:VLDB:2010:NeumannW}.
These indices are intended for optimising access to a single, centrally managed data storage solution.
In this case, accuracy of the index is not an issue as all changes in the data are under control of the storage solution and are reflected in the index immediately. 

More specific to LOD are indices for optimising federated queries~\cite{P:SIGMOD:2009:NeumannW}, on-demand queries on the Web~\cite{P:WWW:2010:HarthHKP} or looking up data sources relevant to particular schema patterns~\cite{J:JWS:2012:KonrathGSS}.
However, most of these approaches do not deal with index accuracy either. 
Their focus is more on how to implement or make use of the index in specific scenarios.

So far, only few publications address the issue of Linked Data index accuracy. 
This only happens in a context where the dynamics of data is explicitly addressed~\cite{P:ESWC:2013:KaeferAUO,P:COLD:2013:DividinoSGG,P:PROFILES:2014:DividinoGSG}, where implementations trade a loss of index accuracy for an efficient and scalable index computation~\cite{J:JWS:2012:KonrathGSS,P:CSWS:2012:GottronP} or where the accuracy of indices is under investigation itself~\cite{P:ESWC:2014:GottronG,P:PROFILES:2014:Gottron,P:CSWS:2012:GottronP}.
These methods used will be reviewed and explained in more detail in Section~\ref{sec:measures}.

In the context of the ``classical'' Web of documents, there is some work on index accuracy of search engines~\cite{J:Nature:1999:LawrenceG}.
Such analysis is performed by comparing the coverage of relevant web documents for specific queries. 
The results of several individual search engines can be compared to the union of their results as well as to a separately obtained collection of relevant documents which serves as ground truth.
Other works attempt to measure age and freshness of an index~\cite{P:SGIMOD:2000:ChoG}.
Incorporating time information into a change prediction allows for more efficient and effective synchronisation plans between Web based data sources.
Yet another direction of research addressed the question of guarantees of freshness of cached copies for web documents~\cite{J:CN:2000:BrewingtonC}.
However, measuring the freshness and accuracy of search indices for Web documents is different from the accuracy of Linked Data indices insofar as the indices are all of the same type, namely mapping keywords to documents.
In the context of Linked Data indices can be of different types and address specific information needs which are directly encoded in the index structure~\cite{P:ESWC:2014:GottronG}.

% ****************************************************************************************
% ****************************************************************************************
%   Section break
% ****************************************************************************************
% ****************************************************************************************

\section{Abstract Index Models for Linked Data}
\label{sec:index}

As indicated in the previous section, there is a wide range of different index models over LOD. 
In this paper we will not look at specific implementations of indices, but rather operate on an abstract level.
Therefore, we now briefly recall a formalisation of abstract index models over Linked Data~\cite{P:ESWC:2014:GottronG}.
This formalisation will serve as basis for a generic and unified definition of accuracy  measures in Section~\ref{sec:measures}.

On the LOD cloud we can assume data items to be in the form of NQuads~\cite{w3cnquads}.
In an NQuad $(s,p,o,c)$ the entries $s$, $p$, and $o$ correspond to the subject, predicate and object of the RDF triple statement.
The entry $c$ provides the context, i.e. the data source on the Web where this information has been published.
%Formally, we consider the sets of all possible URIs $U$, blank nodes $B$ and literals $L$.
%Then in a quad $(s,p,o,c)$ the subject $s \in U \cup B$ can be a URI or a blank node, the predicate $p \in U$ a URI, the object $o \in  U \cup B \cup L$ a URI, a blank node or a literal and the context $c\in U$ a URI. 

Thus, we define an index model for LOD over a data set $R$ of $(s,p,o,c)$ NQuads.
Depending on the application scenario an index will typically not serve to store all information contained in the NQuads.
Rather it will define a derived set $D$ of managed data items which are of interest in the scenario and typically constitute a restriction of the quads to smaller tuples.
Such restrictions can be, for instance, the RDF triples or even single entries, e.g. the subject or the context URIs.
%In this paper we consider two different types of data item sets: (1) the set $D_{\textit{SPO}} := \{  (s,p,o) \mid \exists c : (s,p,o,c) \in R\} $ of full RDF triples and (2) the set $D_{\textit{USU}} :=  \{ s \mid \exists p,o,c : (s,p,o,c) \in R\}$ of all unique subject URIs (USUs). %, which are defined as follows:
%%
%%\begin{eqnarray}
%%D_{\textit{SPO}} & := & \{  (s,p,o) \mid \exists c : (s,p,o,c) \in R\} \\
%%D_{\textit{USU}} & := & \{ s \mid \exists p,o,c : (s,p,o,c) \in R\}
%%\end{eqnarray}
%%
%The data items in $D_{\textit{SPO}}$ are typically used in a context where an index is based on using a part of the quad information (e.g. the object) to look up matching triple statements (e.g. the subject or the predicate).
%$D_{\textit{USU}}$, instead is typically used in index models where information from several statements  (e.g. a set of several RDF types) is used to look up a specific entity URI. 
%
Furthermore, an index model has to define a set $\mathcal{K}$ of key elements which are used to lookup and retrieve data items.
These key elements are used as domain for a selection function $\sigma : \mathcal{K} \rightarrow \mathcal{P}(D)$ to select a subset of the data items in the index.
%In the context of this paper we consider only data structures for which the selection function $\sigma$ is operating solely on information provided by the Linked Data set $R$ and does not make use of external information or additional meta data (e.g. provenance information).

\begin{figure}[btp]
\centering
	\includegraphics[width=\indexwidth]{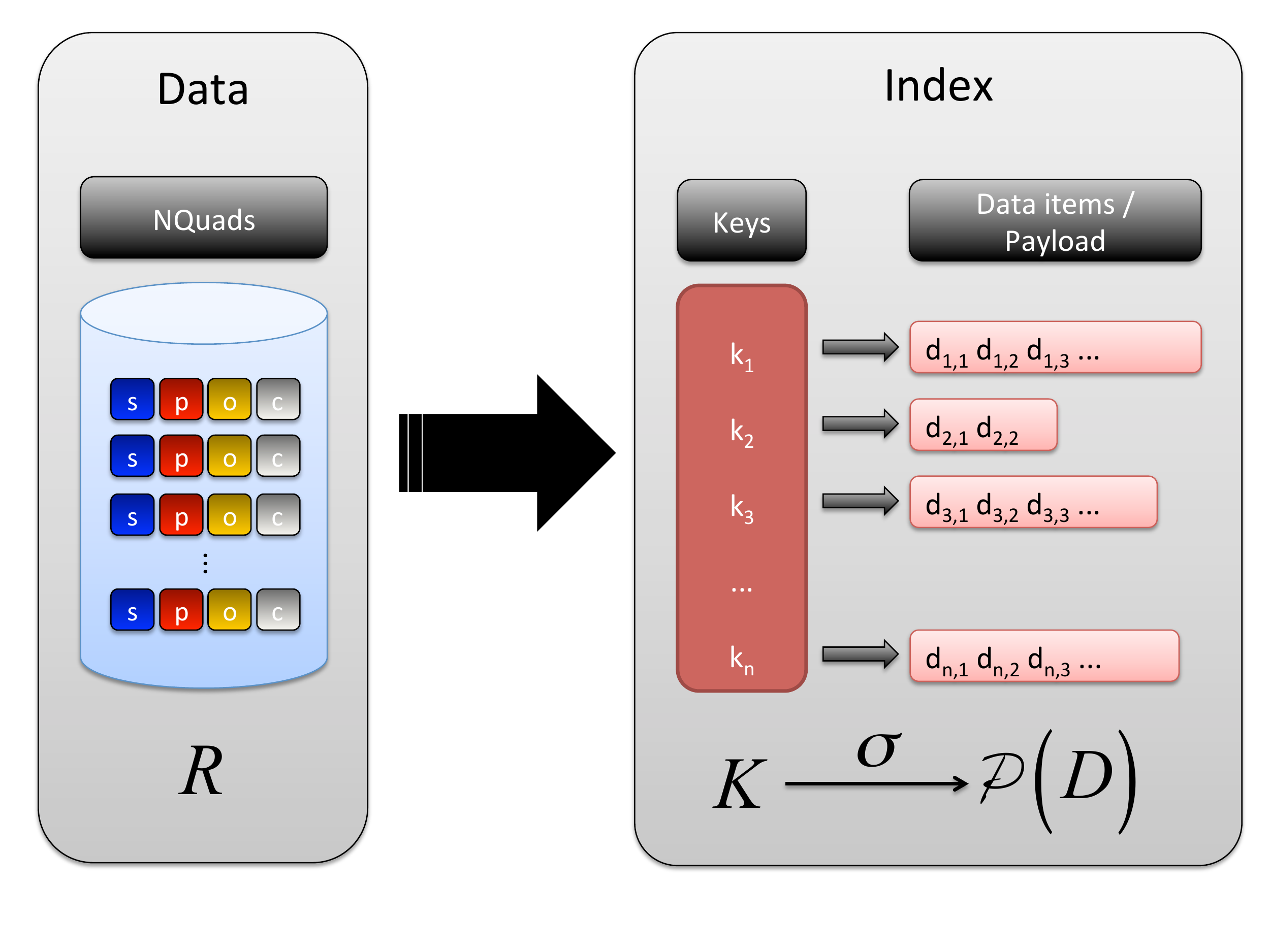} 
\caption{A data set $R$ of NQuads and the elements of an index model over this data.}
\label{fig:abstractindex}
\end{figure}

Eventually, an abstract index model is defined as a tuple $(D,\mathcal{K},\sigma)$ of the stored data items $D$, the key elements $\mathcal{K}$ used for the lookup index and the selection function $\sigma$ to retrieve data from the index.
Figure~\ref{fig:abstractindex} illustrates the elements of an abstract index model $I$ computed over a data set $R$.

% ****************************************************************************************
% ****************************************************************************************
%   Section break
% ****************************************************************************************
% ****************************************************************************************

\section{Measuring Accuracy}
\label{sec:measures}

The formal definition of abstract index models in Section~\ref{sec:index} enables us to now formalise measures for index accuracy in a unified way.
As already indicated in Figure~\ref{fig:concept}, measuring the accuracy of an index will be based on a \emph{perfect} index which is entirely accurate.
In the following, this \emph{gold standard} index will be referred to as $I_\textsc{gs} = (D_\textsc{gs},\mathcal{K}_\textsc{gs},\sigma_\textsc{gs})$ which is built over a data set $R_\textsc{gs}$.
The potentially inaccurate index for which we want to determine its accuracy will be referred to as $I = (D,\mathcal{K},\sigma)$.
Note, that it is not necessary to distinguish between $I$ being inaccurate because it was built over an outdated data set $R$ or because it was computed in an approximative way over the gold standard data set $R_\textsc{gs}$.

The presented measures are divided into four families, based on their methods and underlying ideas for measuring accuracy: (1) index agnostic measures, (2) measures based on the overlap of key elements, (3) distribution based measures and (4) retrieval based measures.

\subsection{Index Agnostic Measures}
\label{subsec:agnostic}

\begin{figure}[btp]
\centering
	\includegraphics[width=\indexwidth]{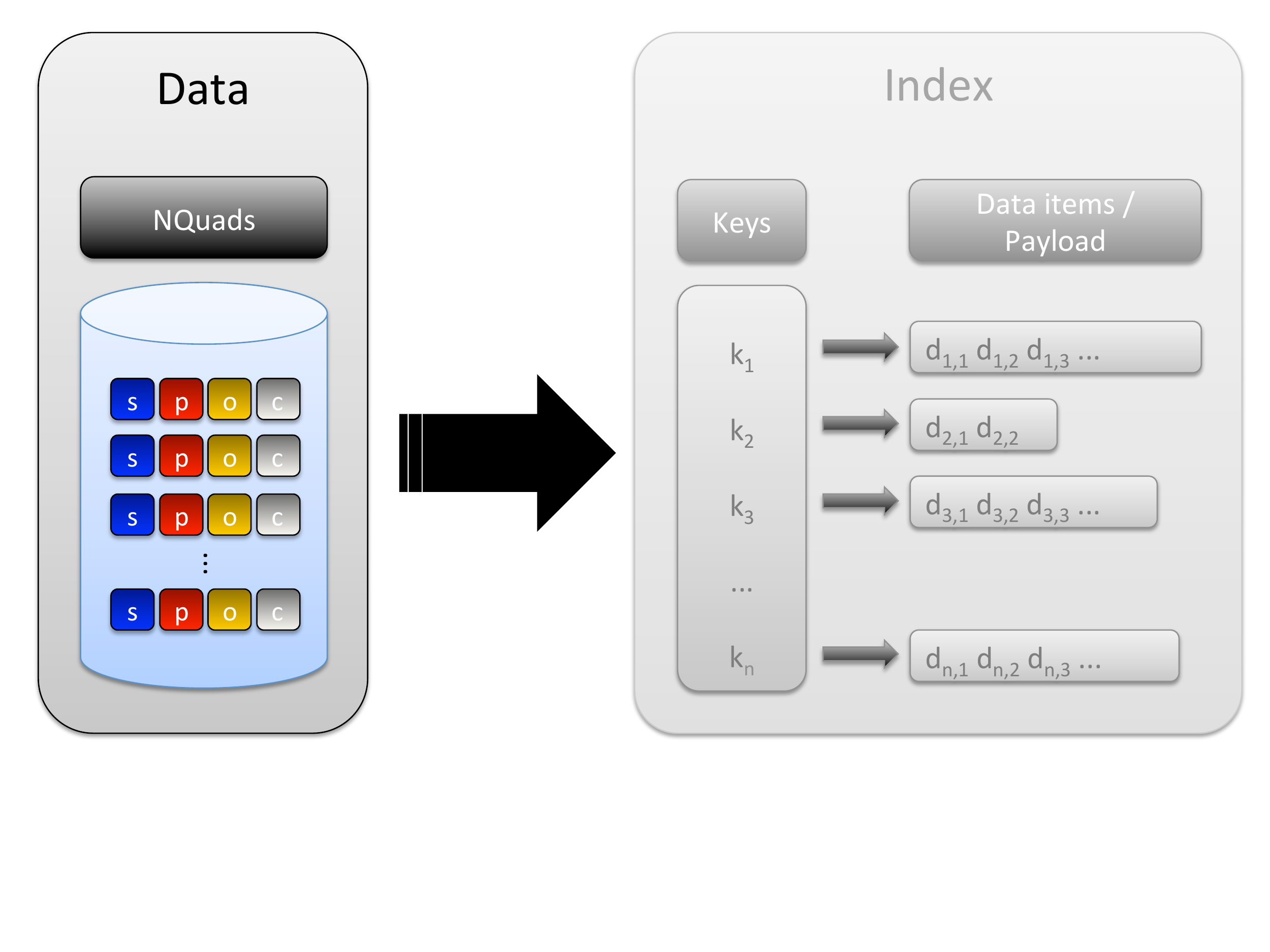} 
\caption{Index agnostic accuracy measures consider only the original data set.}
\label{fig:mAgnostic}
\end{figure}

A direct way to measure the potential impact of data changes on indices is to simply compare the differences in the data itself.
As indicated in Figure~\ref{fig:mAgnostic} such an accuracy measure would accordingly ignore the index and operate only on the original input data.
One established metric for comparing the data sets  is the \emph{Jaccard similarity}.
Applying it to $R_\textsc{gs}$ and $R$, it is defined as:

\begin{equation}
\textit{Jaccard}(R_\textsc{gs},R) = \frac{|R_\textsc{gs} \cap R|}{|R_\textsc{gs} \cup R|}
\end{equation}

A higher similarity value indicates a larger overlap between the data sets while a low value indicates a stronger deviation, i.e. change in the data.

Such an index agnostic measures can always be computed over two versions of a data set at different points in time.
Therefore, it is independent of the index model applied. 
While this might be suitable to measure the evolution of the data in general, it does not make any statement about the impact on specific index models used in specific use case scenarios.
Furthermore, it requires the availability of the full raw data sets and is by design not applicable in settings were indices are computed in an approximative way over the gold standard data set $R_\textsc{gs}$.

\subsection{Overlap of Key Elements}
\label{subsec:overlap}

\begin{figure}[btp]
\centering
	\includegraphics[width=\indexwidth]{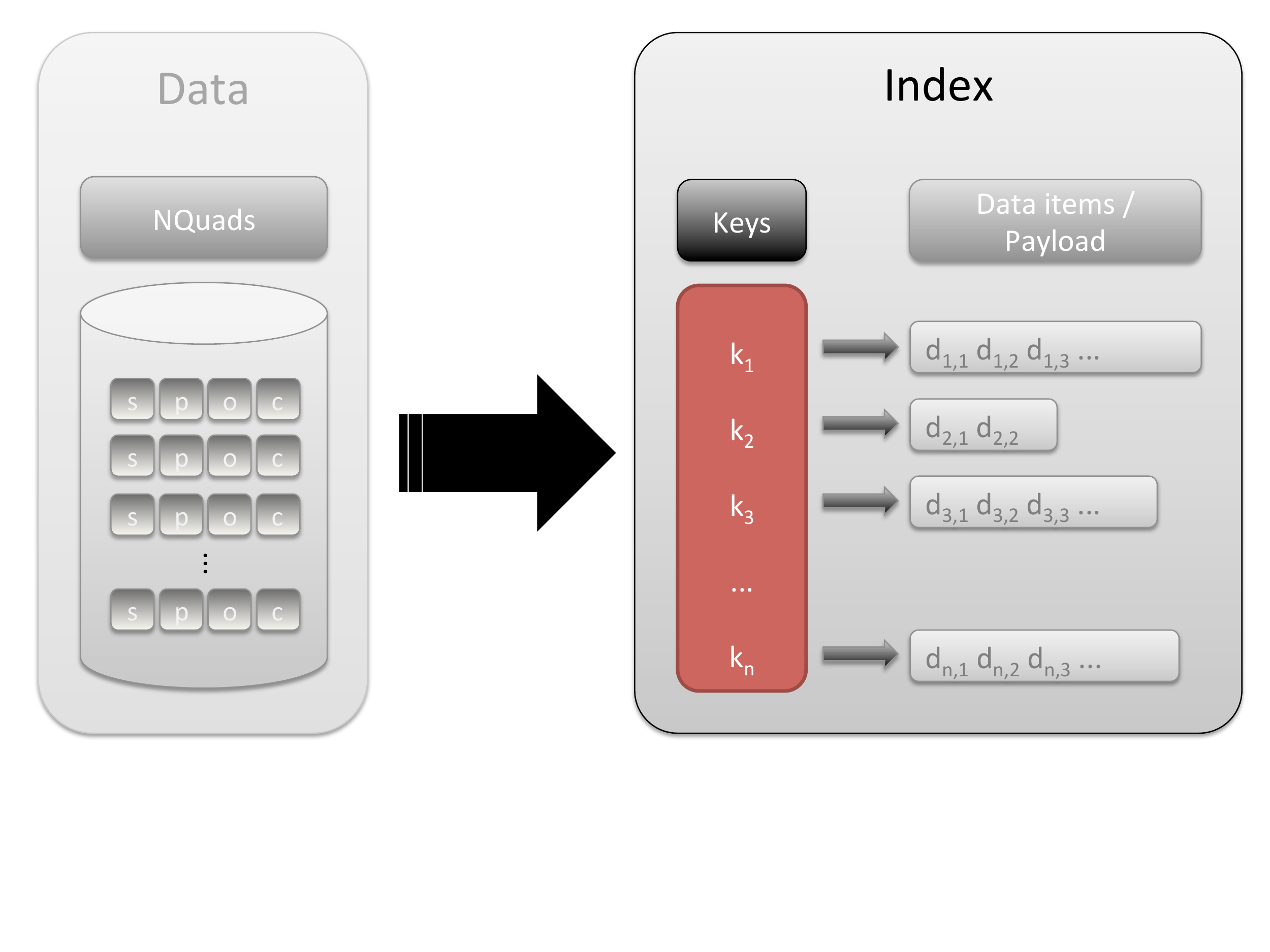} 
\caption{Accuracy measures based on the key elements do not make use of any information about the data items.}
\label{fig:mKeys}
\end{figure}

A relatively simple approach for comparing indices themselves is to look at the key elements (cf. Figure~\ref{fig:mKeys}).
These are the elements used to retrieve information from the index and, thus, provide the primary access point to the contained data.
Hence, the degree of how far an index reflects the right key elements can already give some insights into its accuracy.
Note, that this type of measure will not consider at all the data elements.
In particular, it might assert an index a perfect accuracy even if the index provides wrong results.

Technically, this approach comes down to comparing two sets: $\mathcal{K}_\textsc{gs}$ and $\mathcal{K}$.
There are various measures for comparing sets.
In the context of measuring the accuracy of LOD indices, also here the \emph{Jaccard similarity} has been used~\cite{P:ESWC:2014:GottronG}:

\begin{equation}
\textit{Jaccard}(\mathcal{K}_\textsc{gs},\mathcal{K}) = \frac{|\mathcal{K}_\textsc{gs} \cap \mathcal{K}|}{|\mathcal{K}_\textsc{gs} \cup \mathcal{K}|}
\end{equation}

Again, a higher similarity value indicates a larger overlap between the sets of key elements while a low value indicates a stronger deviation.
In this way we can get an impression of how stable is the set of elements used for indexing in the different indexing approaches.

A further measures to operate on the set of key elements is the asymmetric \emph{recall}, which measures which fraction of the gold standard key set is covered in an inaccurate index.

\begin{equation}
r(\mathcal{K}_\textsc{gs},\mathcal{K}) = \frac{|\mathcal{K}_\textsc{gs} \cap \mathcal{K}|}{|\mathcal{K}_\textsc{gs}|}
\end{equation}

The rational for using this variation is that key elements in an inaccurate index which are not available in the gold standard might not be used to retrieve information as well. However, this rational is debatable as a query with one of this key elements missing in the gold standard will lead to retrieving false positive information from the inaccurate index, a fact which is neglected by recall.

\tg{How to implement: SPARQL ask queries? }

%* Overview
%
%* Definition
%
%* Use cases
%
%
%In addition to the metrics for comparing the density estimates, we use the Jaccard-similarity over the set of key elements.
%Let $\mathcal{K}_1$ and $\mathcal{K}_2$ be the sets of key elements derived at two points in time. 
%Then the Jaccard-similarity is defined as:
%
%\begin{equation}
%\textit{Jaccard}(\mathcal{K}_1,\mathcal{K}_2) = \frac{|\mathcal{K}_1 \cap \mathcal{K}_2|}{|\mathcal{K}_1 \cup \mathcal{K}_2|}
%\end{equation}
%
%A higher similarity value indicates a larger overlap between the sets of key elements while a low value indicates a stronger deviation.
%In this way we can get an impression of how stable the set of elements used for indexing is in the different indexing approaches.
%\cg{gibt es fuer die metriken referenzen?}

\subsection{Distribution Based Measures}
\label{subsec:distribution}

\begin{figure}[btp]
\centering
	\includegraphics[width=\indexwidth]{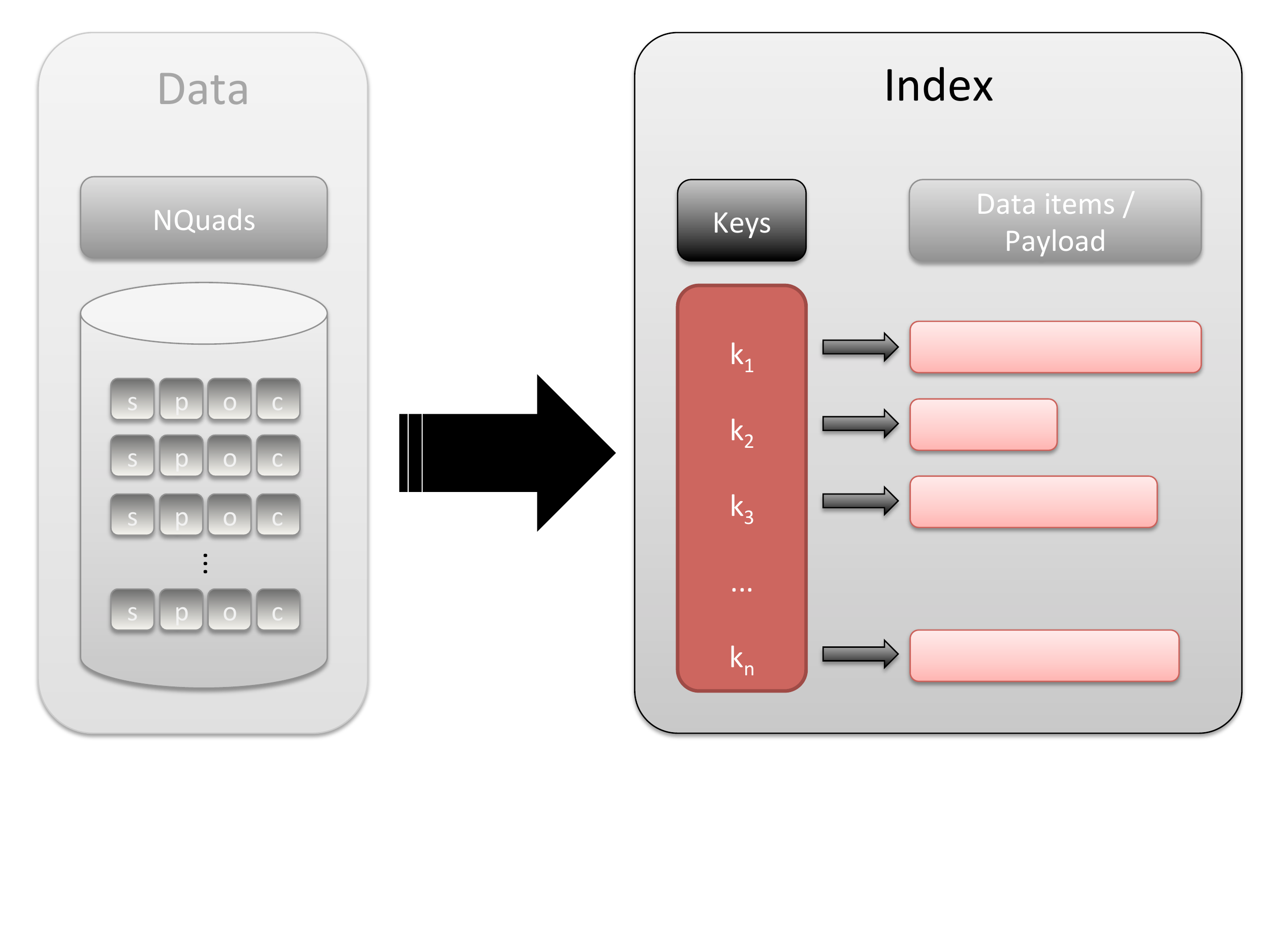} 
\caption{Distribution based measures compare distributions of the data over key elements.}
\label{fig:mDistrib}
\end{figure}

One option to include some more information in the evaluation of index accuracy is pursued by distribution based approaches~\cite{P:ESWC:2014:GottronG,P:PROFILES:2014:Gottron}.
The advantage is that these approaches take into consideration the volume of data items retrieved for a given key element.
This idea is visualised in Figure~\ref{fig:mDistrib} by omitting the concrete data items assigned to a key element.
The volume of the data is used to model probabilistic distributions for which there are several well established and well understood metrics for comparison.
However, they do not distinguish the data items but merely use their count information.
Hence, also here there is a certain risk of asserting a perfect index accuracy while the index actually is not accurate.

Estimating a distribution over an index is based on determining how probable it is for an element to belong to one specific index key  $k$ and---conversely---the amount of data obtained when querying the index for this key element $k$.
If we consider the distribution over an index $I=(D,\mathcal{K},\sigma)$, this effectively corresponds to modelling a random variable $X$ taking values of the key elements $\mathcal{K}$.
The estimated density gives the distribution of this random variable $X$.
This means we determine the probability $P(X=k)$ for each entry $k\in \mathcal{K}$ to be associated with a data item.
%
%\paragraph{Density estimation}
%
To estimate the densities we can use the count information of data elements associated with the key elements in an index.
This corresponds to using a maximum likelihood estimation to derive the probability, i.e.

\begin{equation}
P(X=k) = \frac{|\sigma(k)|}{\sum_{k'\in\mathcal{K}}|\sigma(k')|}
\end{equation}

where $\sigma(k)$ indicates the result set obtained from an index when querying for a specific key element $k$.

%\paragraph{Smooting techniques}

As we consider inaccurate or outdated indices it is likely that the set of key elements does not match (see also Section~\ref{subsec:overlap}).
Thus, it can happen that certain key elements might be available in the perfect gold standard index, but not in an outdated or approximated index.
When comparing densities over indices we need to consider the effect of such zero-size entries.
Using a maximum likelihood estimation for the densities would lead to zero probabilities for certain events which renders comparison of densities impractical.
%We investigate different smoothing techniques for overcoming zero probabilities.
We apply smoothing to overcoming zero probabilities.
We make use of Lidstone smoothing which adds a small constant value of  $\lambda$ to all counts obtained for the number of results $|\sigma(k)|$.
The parameter $\lambda$ is set to $0.5$ in our experiments which has shown to provide good results in prior work~\cite{P:PROFILES:2014:Gottron}.

The main idea of measuring index accuracy on the basis of data distributions is to compare their density function.
Common metrics to compare density functions are cross entropy, Kullback-Leibler divergence and perplexity. %~\cite{J:Bell:1948:Shannon}.
Let us briefly review the definitions of these metrics and explain their interpretation.

Assuming two probability distributions $P(X)$ and $P_{\textsc{gs}}(X)$ for the inaccurate index $I$ and the gold standard index $I_{\textsc{gs}}$.
Then \emph{cross entropy} $H(P_{\textsc{gs}},P)$ is defined as:

\begin{equation}
H(P_{\textsc{gs}},P) = - \sum_{k\in \mathcal{K}\cup\mathcal{K}_{\textsc{gs}}} P_{\textsc{gs}}(X=k) \log (P(X=k))
\end{equation}

In the context of compression theory, cross entropy can be interpreted as the average number of bits needed to encode events following the distribution $P_{\textsc{gs}}$ based on an optimal encoding scheme derived from $P$.
If the two distributions are equivalent, then cross entropy corresponds to the normal entropy $H(P_{\textsc{gs}})$.
The entropy of $P_{\textsc{gs}}$ also provides a lower bound for cross entropy.
Based on this interpretation, the \emph{Kullback-Leibler divergence} gives the deviation in entropy (or overhead in encoding) relative to the entropy for $P_{\textsc{gs}}$ and is defined as:

\begin{equation}
D_{\subtxt{KL}}(P_{\textsc{gs}},P) = H(P_{\textsc{gs}},P) - H(P_{\textsc{gs}}) 
\end{equation}

Therefore, if two distributions are equivalent, they have a Kullback-Leibler divergence of zero. 
%This is a desirable feature for our evaluation as it renders the comparison of distributions of evolving data  independent from the different levels of the entropy observed for different index structures.

\emph{Perplexity}, instead, provides an evaluation of a distribution by giving the number of events (in our case key elements) which under a uniform distribution would yield the same entropy value. 
As such it is considered to be more easily interpretable by humans than the somewhat abstract entropy values.
Perplexity itself is defined over entropy values, though. 
Here we formulate it directly on the basis of cross entropy:

\begin{equation}
\textit{PP}(P_{\textsc{gs}},P) = 2^{H(P_{\textsc{gs}},P)}
\end{equation}

Perplexity is a standard metric for evaluating probabilistic models. 
The lower the perplexity is, the better a model explains observed data and the more truthful are its estimates of the probabilities.
%When looking at  perplexity over the cross entropy $H(P,P_{\textsc{gs}})$ in particular, there is also another interesting interpretation of the values.
%If perplexity is higher than the number of events considered, then using a simple uniform distribution instead of $P_{\textsc{gs}}$ would correspond to a better approximation of the distribution $P$.
Furthermore, the interpretation of perplexity relative to the event space of key elements allows for a normalisation.
The normalised perplexity $\textit{PP}_{\subtxt{norm}}$ is defined as:

\begin{equation}
\textit{PP}_{\subtxt{norm}}(P_{\textsc{gs}},P) = \frac{\textit{PP}(P_{\textsc{gs}},P)}{|\mathcal{K}_{\textsc{gs}}|}
\end{equation}

\subsection{Retrieval Based Measures}

\begin{figure}[btp]
\centering
	\includegraphics[width=\indexwidth]{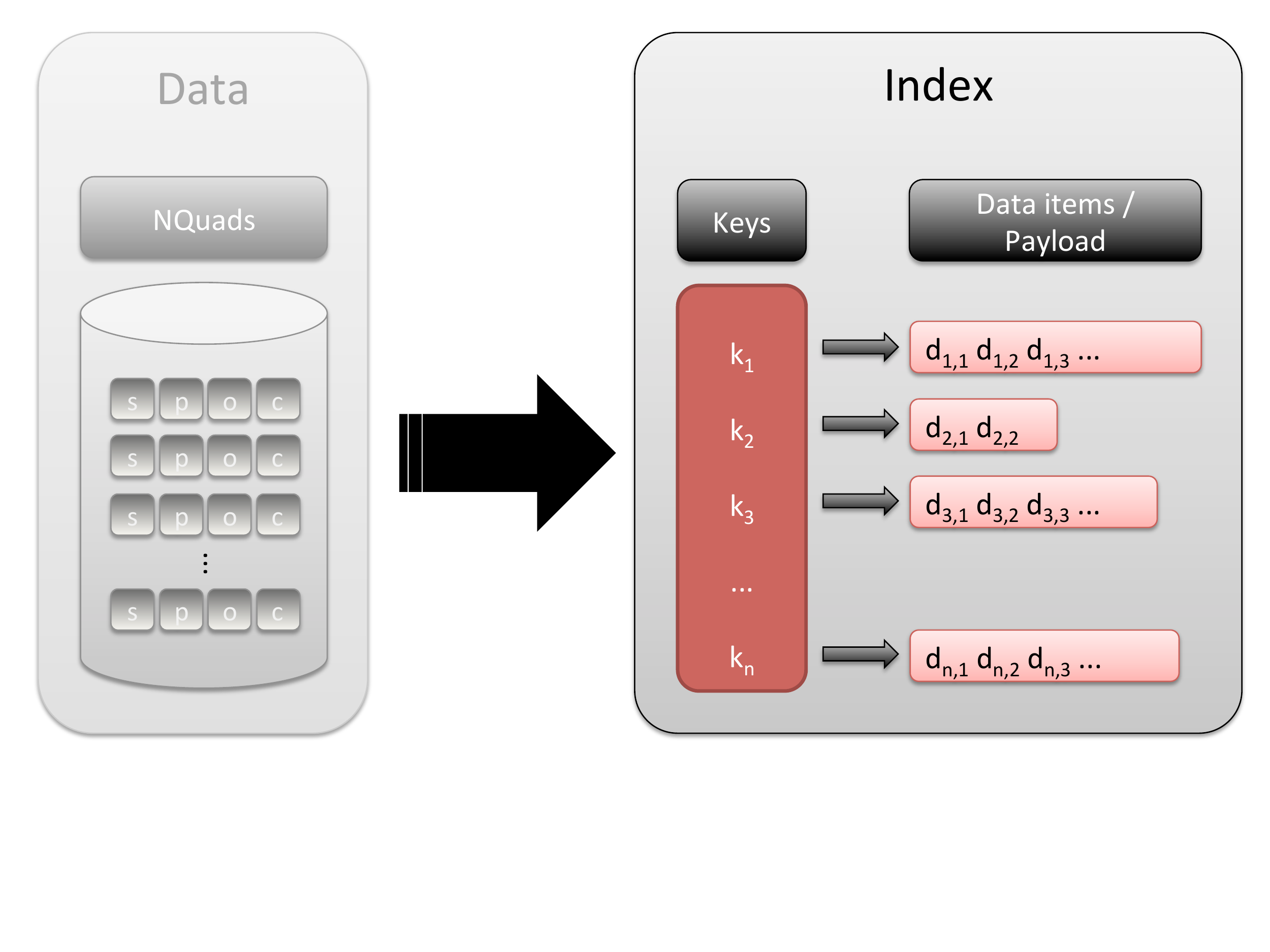} 
\caption{Retrieval based measures include the entire index in the evaluation.}
\label{fig:mRetrieval}
\end{figure}

The third type of measures are directly evaluating the results of queries posed towards an index.
\emph{Precision} and \emph{recall} are two such measures which are typically employed in Information Retrieval.
While in general, this approach can be based on a fixed and predefined set of queries, such a set of queries might not be available in most cases.
Furthermore, a fixed set of queries will always address only a certain fraction of the index.
Thus, an alternative approach which has been followed in related work is to pose virtually all possible queries which could be asked over the gold standard as well as the inaccurate index.
In this way, the measures make use of all information stored in the index as indicated in Figure~\ref{fig:mRetrieval}.

For one specific query, i.e. a key item $k$, precision and recall are defined as follows:

\begin{equation}
p(k) = \frac{|\sigma(k) \cap \sigma_\textsc{gs}(k)|}{|\sigma(k)|}
\end{equation}

\begin{equation}
r(k) = \frac{|\sigma(k) \cap \sigma_\textsc{gs}(k)|}{|\sigma_\textsc{gs}(k)|}
\end{equation}

There is, however a distinction to be made with respect to the way of aggregating results over the set of all queries.
Namely, there are micro- and macro-averages which can be computed.

Macro-average builds an average over all precision or recall values for all of the queries.
Formally this gives (here for the example of precision):

\begin{equation}
p_\subtxt{macro} = \frac{1}{\left | \mathcal{K} \cup \mathcal{K}_{\textsc{gs}} \right |} \sum_{k \in \mathcal{K} \cup \mathcal{K}_{\textsc{gs}}} \frac{|\sigma(k) \cap \sigma_\textsc{gs}(k)|}{|\sigma(k)|}
\end{equation}

Micro-average, instead, aggregates the set sizes of the observed result sets and their intersection.
Precision and recall are then computed on these aggregated count information.
Formally, it is defined as:

\begin{equation}
p_\subtxt{micro} = \frac{ \sum_{k \in \mathcal{K}\cup\mathcal{K}_{\textsc{gs}}} |\sigma(k) \cap \sigma_\textsc{gs}(k)|}{\sum_{k \in \mathcal{K}\cup\mathcal{K}_{\textsc{gs}}} |\sigma(k)|}
\end{equation}

The difference of the two ways for computing averages can be seen as follows.
Macro-average gives equal weight to each query. 
Hence a lot of good or bad performing queries which provide small result sets will strongly affect the overall score.
Computing a micro-average, instead, considers also the size of the results set in the overall aggregated score.
Larger result sets and their performance have a stronger impact on the overall result.

% ****************************************************************************************
% ****************************************************************************************
%   Section break
% ****************************************************************************************
% ****************************************************************************************

\section{Theoretic Analysis of Measures}
\label{sec:theory}

Let us consider the measures introduced above in Section~\ref{sec:measures} and compare them on a theoretic level.
To this end we will consider the following four  criteria and argue for and against the individual approches, independent of a concrete scenario, index model or data set.

\begin{description}
\item[Sensitivity to Data Changes] 
Not all measures perfectly reflect the accuracy of an index. 
In some cases a change in the data and a loss of accuracy of an outdated index will not be captured by an accuracy measure.
The higher the sensitivity of a measure the more precise and reliable it is in evaluating an index's ability to provide truthful results.
In the context of this paper the theoretic sensitivity is judged in a relative and categorial way, by assigning the measures a low (\tickBad), medium (\tickOk) or perfect (\tickGood) sensitivity. 
A perfect sensitivity implies that each error in an inaccurate index will be detected.

\item[Data Volume] 
When computing the measures, one question is how much data needs to be obtained from a perfect index at the original data source. 
A smaller volume of data which needs to be transfered from separate indices makes it easier to compute the measures remotely and only request the needed information from the actual computing nodes running the index.
Furthermore, the computational complexity of the presented measures is linear in the amount of data items to consider.
Thus, the data volume indicates how long the computation of the measures will take.
For this reason, the data volume will be given in big $O$ notation based on the elements of abstract index models.

\item[Normalised Value Range] 
A normalised value range is favourable in an index accuracy measure.
If a measure provides a normalised value range it is easier to compare its values across experiments. 
This can affect both: the comparison of accuracy of the same index over different data sets as well as the comparison of different indices over the same data set.
Having a normalised value range is a binary feature of an accuracy measure, i.e. a measure can either have a normalised range (\tickYes) or not (\tickNo).

\item[Index Agnostic] If a measure is index agnostic it can be computed independently of a concrete index model or implementation. 
The advantage is, that the computation of one single value over the data is sufficient to judge a (potential) impact on different types of indices.
A conceptual disadvantage is, that index agnostic measures cannot be used to evaluate approaches performing an approximate index computation.
Also being index agnostic is a binary feature, i.e. the measure is index agnostic (\tickYes) or not (\tickNo).

%\item[Symmetric] Measures might be symmetric, which means it does not matter from which perspective they are evaluated (which data set to use as ground truth).
%Asymmetric measures make a distinction.
\end{description}

Table~\ref{tab:theory} summarises the theoretic evaluation of the measures from Section~\ref{sec:measures} w.r.t the above mentioned criteria.
Within each family of measures, the characteristics are mostly the same.
Typically, the measures have the same sensitivity, require the same volume of data and are all either index specific or index agnostic.

Judging from this theoretic point of view, the retrieval based measures seem favourable.
After all they directly address the purpose of an index and measure the accuracy using the core functionality.
Furthermore, they provide normalised values which renders them suitable for comparisons across data sets and types of indices.
However, they require a large volume of data. 
Basically, all information of the entire index has to be considered for a comparison, which---depending on the type of index---might be as costly as comparing the entire underlying RDF data set.
Under this aspect, the measures using the overlap of key elements and the distribution based measures need less information.
They merely operate on the set of key elements and---in the case of the distribution based measures---a size estimation of how many elements are associated with a key element.
Such size estimations can be realised using count operators over the existing indices, causing only a slight computational overhead.
Their drawback is that they might not always capture perfectly the accuracy of an index.

\begin{table}[tb]
\centering 
\caption{Theoretic analysis of measures for index accuracy}
\label{tab:theory}
\resizebox{\columnwidth}{!}{%
\begin{tabular}{l | c c c c }
\toprule
Measure						& Sensitivity	& Normalised range & Volume 		& Index agnostic\\
\midrule
{\bf Index agnostic measures} \\
Jaccard (RDF triples) 		& \tickGood	& \tickYes	& $O(|R| + |R_\textsc{gs}|)$  & \tickYes	\\
\midrule
{\bf Overlap of key elements} \\
Jaccard (Key elements) 		& \tickBad	& \tickYes	& $O(|\mathcal{K} | + | \mathcal{K}_\textsc{gs}|)$ & \tickNo	 \\
Recall (Key elements) 		& \tickBad	& \tickYes	& $O(|\mathcal{K} | + | \mathcal{K}_\textsc{gs}|)$ & \tickNo	 \\
\midrule
{\bf Distribution based measures} \\
Cross-Entropy				& \tickOk	& \tickNo	& $O(|\mathcal{K} | + | \mathcal{K}_\textsc{gs}|)$  & \tickNo	\\
KL-Divergence 				& \tickOk	& \tickNo	& $O(|\mathcal{K} | + | \mathcal{K}_\textsc{gs}|)$  & \tickNo	\\
Perplexity					& \tickOk	& \tickNo	& $O(|\mathcal{K} | + | \mathcal{K}_\textsc{gs}|)$  & \tickNo	\\
Normalised Perplexity		& \tickOk	& \tickYes	& $O(|\mathcal{K} | + | \mathcal{K}_\textsc{gs}|)$  & \tickNo	\\
\midrule
{\bf Retrieval based measures} \\
Recall (macro-avg)			& \tickGood	& \tickYes	& $O(|\mathcal{D} | + | \mathcal{D}_\textsc{gs}|)$  & \tickNo	\\
Precision (macro-avg)		& \tickGood	& \tickYes	& $O(|\mathcal{D} | + | \mathcal{D}_\textsc{gs}|)$  & \tickNo	\\
Recall (micro-avg)			& \tickGood	& \tickYes	& $O(|\mathcal{D} | + | \mathcal{D}_\textsc{gs}|)$  & \tickNo	\\
Precision (micro-avg)		& \tickGood	& \tickYes	& $O(|\mathcal{D} | + | \mathcal{D}_\textsc{gs}|)$  & \tickNo	\\
\bottomrule
\end{tabular}%
}
\end{table}

% ****************************************************************************************
% ****************************************************************************************
%   Section break
% ****************************************************************************************
% ****************************************************************************************

\section{Empirical Comparison of Measures}
\label{sec:empirical}

Following the theoretical evaluation of the measures we now consider how they perform in practice.
The question we want to answer is which measures are closely correlated to each other in practice.
The aim is to find out which measures seem to be redundant and do not need to be compared.
This might help in reducing overhead in practical evaluation and in identifying efficient measures to approximate or even substitute more complex measures.
In particular, a correlation analysis can also address the question how large is the effect of data changes in practice on the less sensitive key element and distribution based measures, i.e. how far they still agree with the measures having a perfect sensitivity.

To perform such an evaluation we need a set of indices (for the index specific measures) and a data set which undergoes realistic changes.
As data set we can employ the weekly snapshots of a well defined part of the Linked Data cloud provided by the Dynamic Linked Data Observatory (DyLDO)~\cite{P:ESWC:2013:KaeferAUO}.
The range of index models is wide and it is beyond the scope of this paper to evaluate all types of index models.
Thus, let us chose a few selected index models which cover a good range of index granularities and scope.

The indices are taken vom related work and operate on different types of data they return: (1) full triple statements ($D_{\subtxt{triple}}$), (2) single URIs ($D_{\subtxt{URI}}$) or (3) solely the source where relevant information can be found on the Linked Data cloud ($D_{\subtxt{context}}$).
In the following, the index models are described briefly along with a formal definition for clarity.
More details can be found in original publications.

\noindent\begin{samepage}
{\bf Subject Index:} This type of index simply uses URIs as key elements and its selection function returns all triple statement containing a given URI in the subject position.
Using the abstract notation introduced in Section~\ref{sec:index}, such an index  can be formalised as $I_{\subtxt{S}} := (D_{\subtxt{triple}}, \mathcal{K}_{\subtxt{S}}, \sigma_{\subtxt{S}})$ where:  
\vspace{-2mm}
\begin{itemize}
\item Key elements: $\mathcal{K}_{\subtxt{S}} := \{ s \in U\mid \exists p,o,c : (s,p,o,c) \in R\}$
\item Selection function: $\sigma_{\subtxt{S}}(k) := \{(s,p,o) \mid s = k \}$
\end{itemize}
\end{samepage}

\noindent\begin{samepage}
{\bf RDF Type Index:} Making use of the specific semantics behind \code{rdf:type} statements, this index is used to look up all entities (i.e. URIs) which are of a particular RDF class type~\cite{P:ESWC:2014:GottronG}.
Formally it can be defined as $I_{\subtxt{T}} := (D_{\subtxt{URI}}, \mathcal{K}_{\subtxt{T}}, \sigma_{\subtxt{T}})$, with:  
\vspace{-2mm}
\begin{itemize}
\item Key elements: $\mathcal{K}_{\subtxt{T}} := \{ o\mid \exists s,c : (s,\code{rdf:type},o,c) \in R\} \cup \allowbreak \{ s\mid \exists c : (s,\code{rdf:type},\allowbreak\code{rdfs:Class},c) \in R\}$ 
\item Selection function: $\sigma_{\subtxt{T}}(k) := \{s \mid \exists c : (s,\code{rdf:type},k,c) \in R \}$
\end{itemize}
\end{samepage}

\noindent\begin{samepage}
{\bf RDF Type Set (TS) Index:} Extending the RDF type class index, this index gives all entities for a specific set of class types, i.e. all entities which satisfy being of exactly all types given in the set (and of no other type)~\cite{J:JWS:2012:KonrathGSS}.
The formal definition is given by $I_{\subtxt{TS}} := (D_{\subtxt{URI}}, \mathcal{K}_{\subtxt{TS}}, \sigma_{\subtxt{TS}})$, where: 
\vspace{-2mm}
 \begin{itemize}
\item Key elements: $\mathcal{K}_{\subtxt{TS}} := \mathcal{P}(\mathcal{K}_{\subtxt{T}})$
\item Selection function: $\sigma_{\subtxt{TS}}(k) := \{s \mid (\forall t \in k : (\exists c : (s,\code{rdf:type},t,c) \in R)) \wedge \forall (s,\code{rdf:type},o,c) \in R : (o \in k)\}$
\end{itemize}
\end{samepage}

\noindent\begin{samepage}
{\bf Property Set (PS) Index:} Using sets of predicates---also known as characteristics sets---this index retrieves entities based on the properties with which they have been described in RDF~\cite{P:SIGMOD:2009:NeumannW}.
Formally it can be defined as $I_{\subtxt{PS}} := (D_{\subtxt{URI}}, \mathcal{K}_{\subtxt{PS}}, \sigma_{\subtxt{PS}})$, with:  
\vspace{-2mm}
\begin{itemize}
\item Key elements: $\mathcal{K}_{\subtxt{PS}} := \mathcal{P}(\mathcal{K}_{p})$
\item Selection function: $\sigma_{\subtxt{PS}}(k) := \{s \mid (\forall p \in k : (\exists o,c : (s,p,o,c) \in R)) \wedge \forall (s,p,o,c) \in R : (p \in k)\}$
\end{itemize}
\end{samepage}

\noindent\begin{samepage}
{\bf Extended Characteristic Set (ECS) Index:} This index essentially combines the TS and PS index and characterises Linked Data entities by both: their types and properties~\cite{P:COLD:2013:DividinoSGG}.
Formally it is defined by $I_{\subtxt{ECS}} := (D_{\subtxt{URI}}, \mathcal{K}_{\subtxt{ECS}}, \sigma_{\subtxt{ECS}})$, where: 
\vspace{-2mm}
\begin{itemize}
\item Key elements: $\mathcal{K}_{\subtxt{ECS}} := \mathcal{P}(\mathcal{K}_{\subtxt{P}}\cup \mathcal{K}_{\subtxt{T}})$
% \item Selection function: $\sigma(x)_{ECS} := \{s : \forall p \in x \cap \mathcal{K}_{p} : (\exists (s,p,o) \in R) \wedge \forall (s,p,o) \in R : (p \in x \cap \mathcal{K}_{p}) \wedge \forall t \in x \cap \mathcal{K}_{T} : (\exists (s,\code{rdf:type},t) \in R) \wedge \forall (s,\code{rdf:type},t),o) \in R : (o \in x \cap \mathcal{K}_{T}) \wedge \}$
\item Selection function: $\sigma_{\subtxt{ECS}}(k) := \{s \mid ( \forall p \in k \cap \mathcal{K}_{\subtxt{P}} : (s \in \sigma_{\subtxt{PS}}(p) )) \wedge (\forall t \in k \cap \mathcal{K}_{\subtxt{T}} : (s \in \sigma_{\subtxt{TS}}(t))) \}$
\end{itemize}
\end{samepage}

\noindent\begin{samepage}
{\bf SchemEX Index over data sources:} The SchemEX index combines induced schematic information about the types of subject URIs, their properties and the types of object URIs they are linked to~\cite{J:JWS:2012:KonrathGSS}. 
To this end it re-uses the concepts of type clusters for the domain and range of observed property sets.
Those concepts are combined using a restricted bi-simulation over property sets and a stratification of the entities in equivalence classes of type sets.
Its formal definition is given by $I_{\subtxt{SchemEX}} := (D_{\subtxt{context}}, \mathcal{K}_{\subtxt{SchemEX}}, \sigma_{\subtxt{SchemEX}})$, where: 
\vspace{-2mm}
\begin{itemize}
%\item \tg{Todo formalization ... Maybe in appendix , because complicated}
\item Key elements: $\mathcal{K}_{\subtxt{SchemEX}} := \mathcal{P}( \mathcal{K}_{\subtxt{TS}} \times \mathcal{P}( \mathcal{K}_{\subtxt{PS}} \times \mathcal{K}_{\subtxt{TS}}))$
\item Selection function: $\sigma_{\subtxt{SchemEX}} (k=(\textit{ts},E)) := \{c \mid \exists s \in \sigma_{\subtxt{TS}}(\textit{ts}) \wedge \forall (\textit{ps},\textit{ts}_{2}) \in E : (s \in \sigma_{\subtxt{PS}}(\textit{ps}) \wedge \exists o \in \sigma_{\subtxt{TS}}(\textit{ts}_2) : (\forall p \in \textit{ps}: ((s,p,o,c) \in R) )\}$
\end{itemize}
\end{samepage}

\begin{figure}[btp]
\centering
  \subfigure[Subject index]{
    \label{sfig:s}    
	\includegraphics[width=\corrwidth]{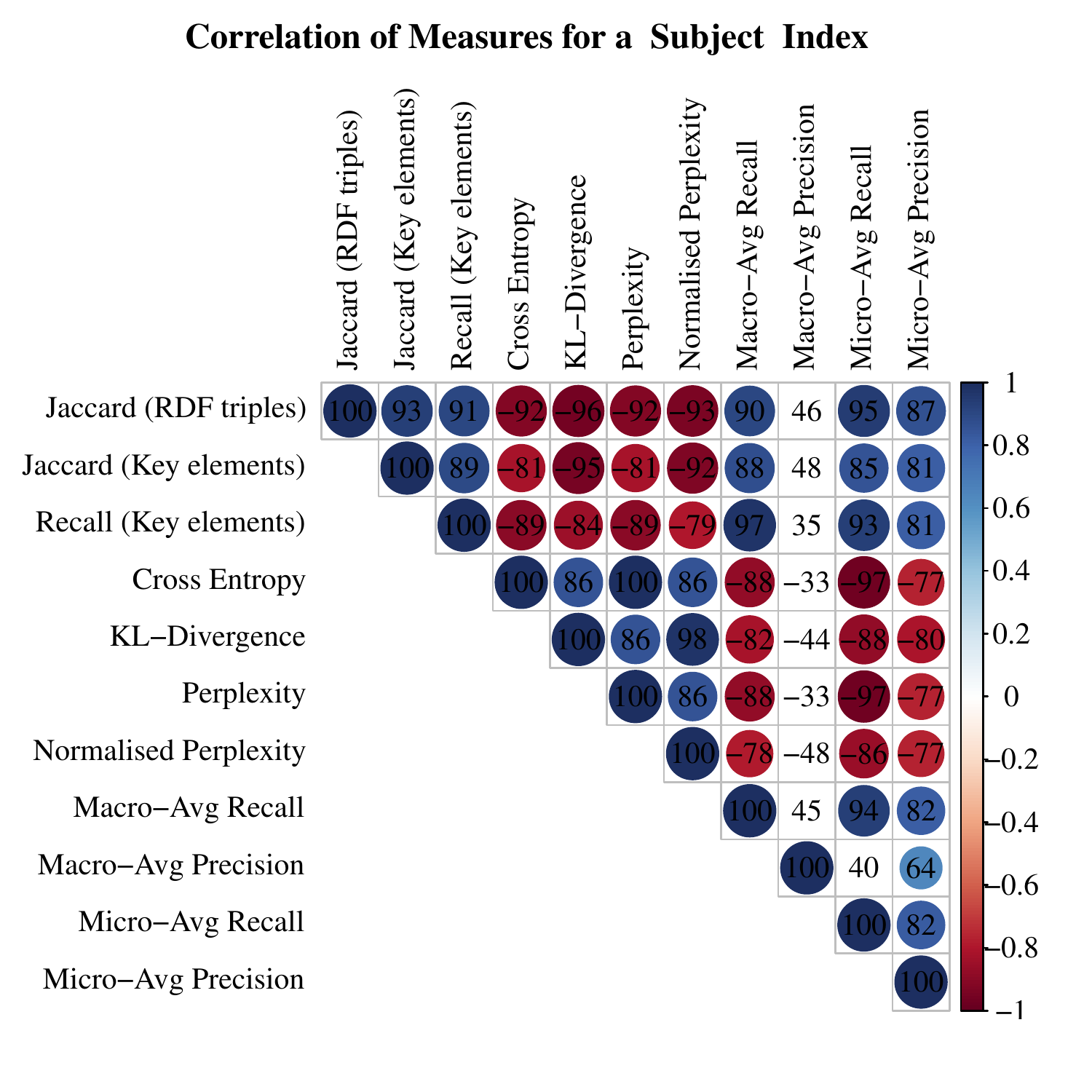} 
  }
  \subfigure[RDF type index]{
    \label{sfig:type}
	\includegraphics[width=\corrwidth]{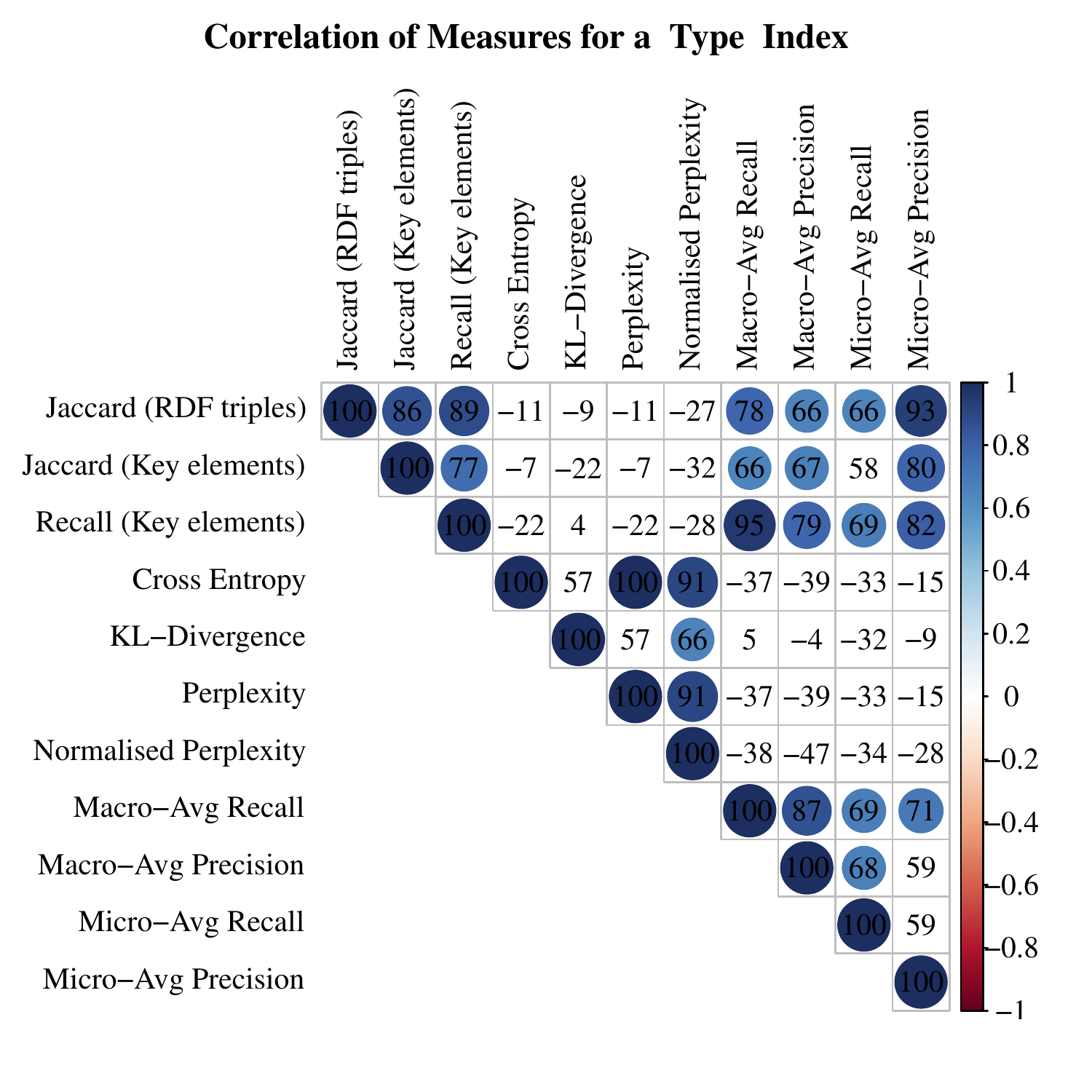} 
  }  
  \subfigure[Type set index]{
    \label{sfig:ts}    
	\includegraphics[width=\corrwidth]{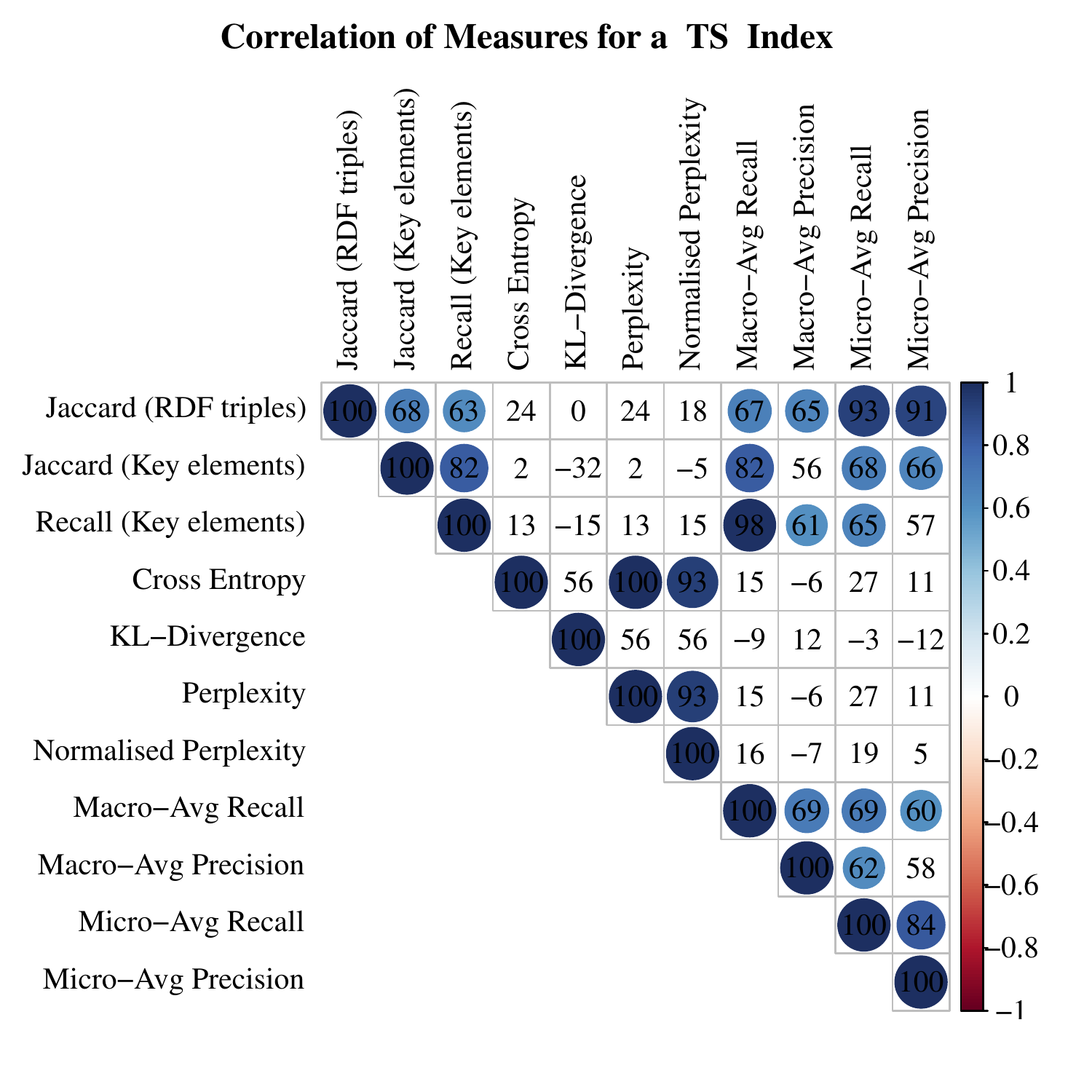} 
  }
  \subfigure[Property set index]{
    \label{sfig:ps}
	\includegraphics[width=\corrwidth]{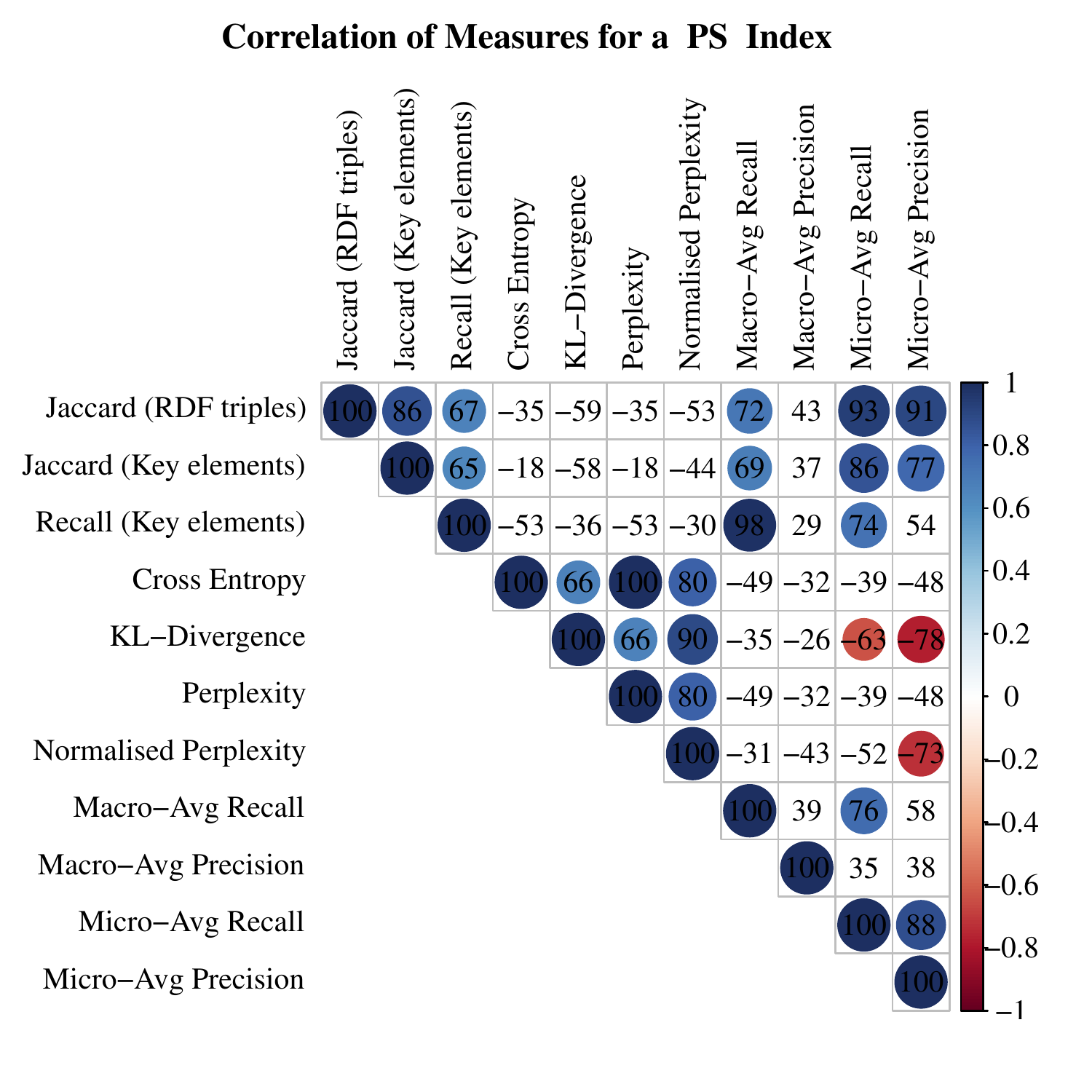} 
  }  
  \subfigure[Extended characteristic set index]{
    \label{sfig:ecs}    
	\includegraphics[width=\corrwidth]{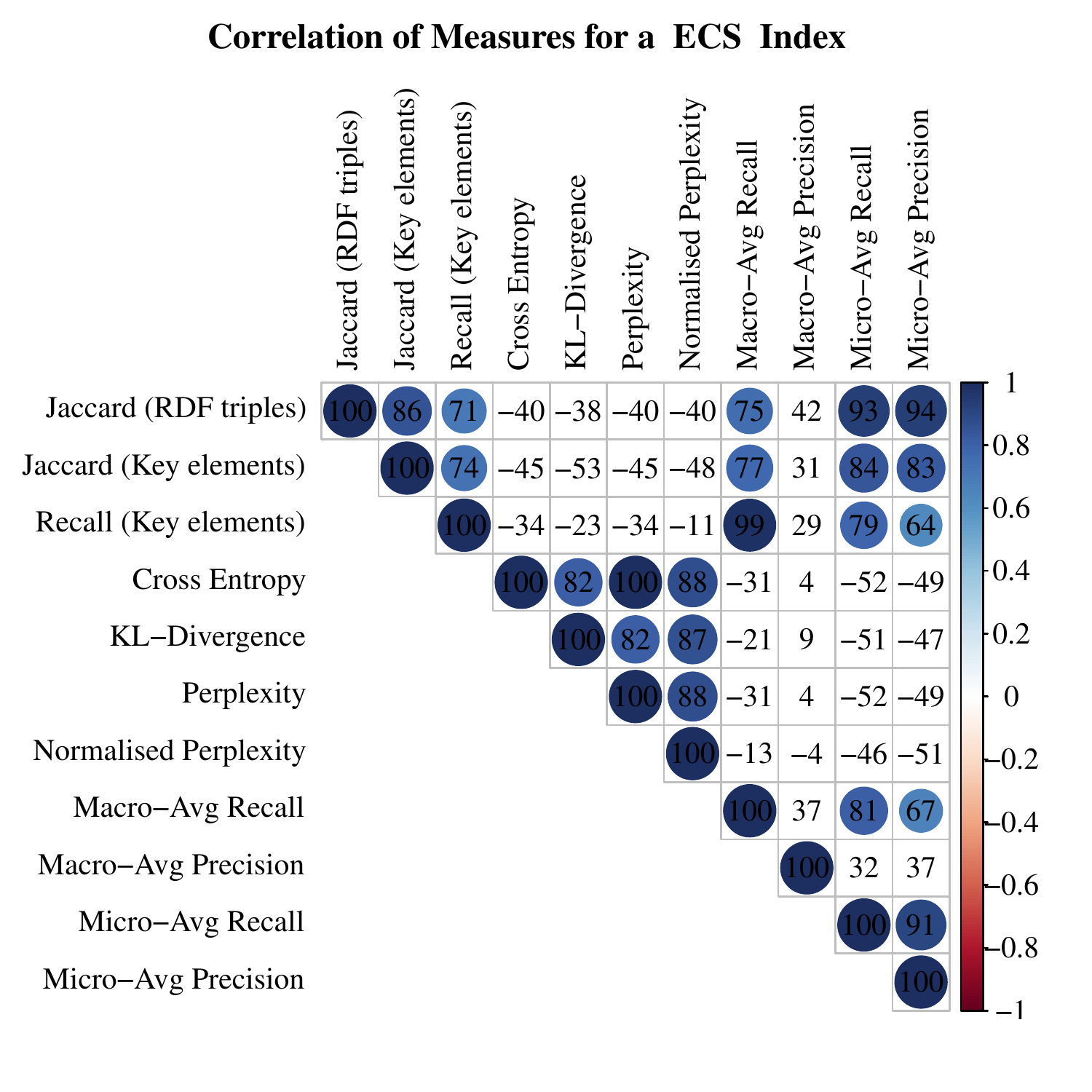} 
  }
  \subfigure[SchemEX index]{
    \label{sfig:schemex}
	\includegraphics[width=\corrwidth]{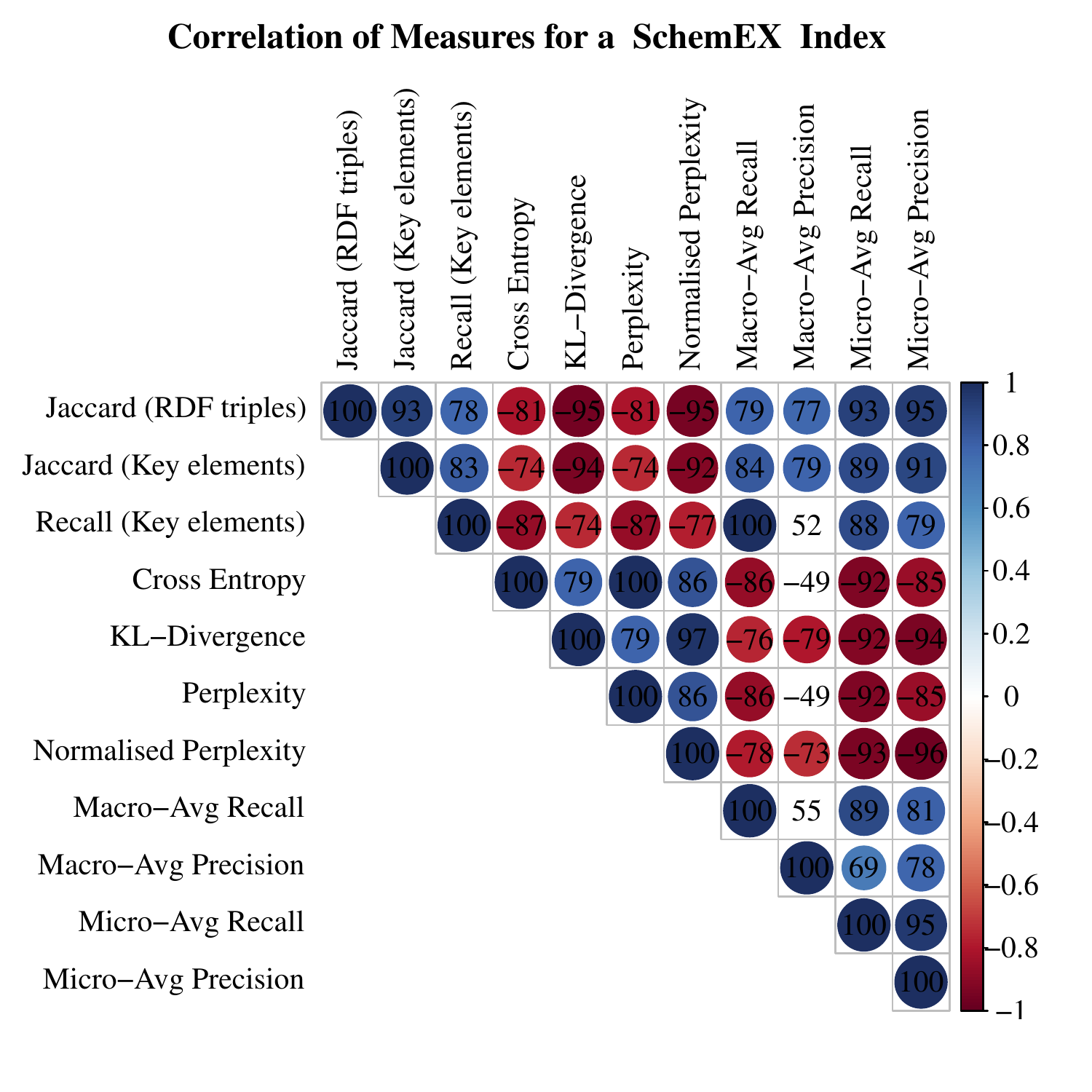} 
  }  
\caption{Correlation of accuracy measures on different indices over an evolving data set.}
\label{fig:correl}
\end{figure}

For the empirical evaluation we computed these indices over 77 snapshots provided by the DyLDO data set.
Assuming an index has been computed over the initial snapshot, we evaluate it accuracy compared to the later on versions of the data sets.
This corresponds to fixing an index over the initial data set and comparing it to gold standard indices over later data sets.
All measures have been computed for all of the evaluated index models.
This gives us for each update of the data set one observation of the accuracy measures.
The results of a spearman rank correlation analysis between all pairs of measures over this data is visualised in Figure~\ref{fig:correl}.

There are several interesting observations to be made.
First of all we observe two index models where the correlations are much stronger in general: the subject index and SchemEX.
All other analysed index models exhibit lower correlation values.
The explanation for this behaviour is the granularity of the indices.
Both, the subject index and SchemEX, generate a relatively large number of key elements over the analysed data set.
Furthermore, they have many key elements in the index which provide only one element entry upon request.
This means on the one hand, that measuring the accuracy of a specific query is often reduced to a binary task: either the index provides the correct element or not.
Therefore, the retrieval based and key element based measures are strongly correlated in a positive way. 
On the other hand, as the change of a data element being assigned to a different key element is typically caused by a single triple, also the Jaccard index over triples correlates strongly with the measures.
Also the changes in the distribution are affected by this.
Changes for key elements with one entry typically leads to a strong shift in the overall distribution, reducing a MLE probability to zero\footnote{Effectively the probability has a small value, slightly higher than zero due to the use of Lidstone smoothing as explained in Section~\ref{subsec:distribution}.}.
This shift in probability mass is recognised by the distribution based measures.
The observation that the correlation here is highly negative comes from the fact that for those measures a high value indicates a strong change, while the other measures use high values to indicate no or only weak changes.

For all other analysed index models, we have a different overall behaviour.
Typically, the measures within each family are correlated (though to a lower extent than for the subject index or SchemEX).
Furthermore, the key element based measures correlate strongly with the retrieval based measures and the Jaccard index over the sets of RDF triples.
The distribution based measures, instead seem to measure something else.
Beyond a few exception, no strong correlation can be observed here. 

Looking at individual measures we can also observe some interesting behaviour:
Across all types of index models macro-average precision seems to be not or only weakly correlated to most other measures.
Hence, the different way of aggregating precision seems to have a strong impact.
The micro-average show a better sensitivity of the accuracy of queries with large results sets.
This seems to make a distinction with other evaluation measures over most analysed index models.
The perfect Spearman correlation between Cross-Entropy and Perplexity, instead,  is conceptual as Perplexity is only a rank-preserving non-linear scaling of Cross-Entropy.

\section{Discussion}
\label{sec:discussion}

Given the analysis in the previous two section, we can observe that a look at the key elements might already give a good indication for the accuracy of an index.
While from a theoretic point of view they might not be very sensitive to changes they correlated well with other, more sensitive measures in the empiric evaluation.
In particular they reflect quite well the retrieval based accuracy measures.
This observations holds for all analysed index models.

Additionally, in many cases the distribution based measures provide further insights.
Given that the volume of data needed to compute these measures is of the same order as for the key elements, such an analysis causes little overhead.

In general, the empirical observations in the previous section were made only over a single data set.
However, given that this data set has been carefully designed to give a representative excerpt of the LOD cloud, we might want to conjecture that density based measures as well as the measures based on key elements suffice to judge the quality of Linked Data indices.
As information on the distributions over key elements in Linked Data index structures is sufficient to compute both types of measures, they are also attractive from a computational point of view.

\section{Summary and Conclusions}
\label{sec:summary}

In this paper I investigated the question of how to measure the accuracy of index structures and data caches which are built in an approximative way or over evolving Linked Data sets.
I gave an overview of different approaches for such measures, compared them on a theoretical and empirical level.
One observation is that  information about the data distribution over key elements seems to provide good insights into index accuracy without the need to access all data in an index.

This observation also motivates a roadmap for future work. 
Operating on samples of Linked Data can provide good estimates of these distributions~\cite{P:PROFILES:2014:Gottron}.
Accordingly an interesting question is how suitable are sampling based strategies for evaluating index accuracy.
This means, that instead of asking all possible queries, we consider only a random sample of queries to obtain distributions.
This might speed up computation or can at least provide certain confidence intervals about what the true accuracy of an index might be.

\vspace{1em} % Remove if space is needed

%\paragraph{Acknowledgements}
%The research leading to these results has received funding from the European Community's Seventh Framework Programme (FP7/2007-2013), REVEAL (Grant agree number 610928).

\bibliography{btw-eval}

\end{document}